\documentclass[12pt]{article}
\usepackage{amsmath}
\usepackage{epic}

\makeatletter
\renewcommand{\@makecaption}[2]{
  \vskip\abovecaptionskip
  \sbox\@tempboxa{\small\sf #1: #2}%
  \ifdim \wd\@tempboxa >\hsize
  \small\sf #1: #2\par
  \else
    \global \@minipagefalse
    \hb@xt@\hsize{\hfil\box\@tempboxa\hfil}%
  \fi
  \vskip\belowcaptionskip}
\makeatother

\newcommand{\preprint}[1]{\rule{0pt}{8pt} \scriptsize #1}

\numberwithin{equation}{section}

\newcommand{\tr}{\operatorname{tr}}

\date{}
\title{Electroweak symmetry breaking from dimensional deconstruction}

\author{Nima Arkani-Hamed\thanks{{\tt
      arkani@bose.harvard.edu}\hfil\break
    Permanent address: \small\sl Department of Physics, UC Berkeley,
    CA 94720}, \ Andrew G. Cohen\thanks{{\tt
      cohen@andy.bu.edu}\hfil\break
    Permanent address: \small\sl Physics Department, Boston
    University, Boston, MA 02215} \ and\ 
  Howard Georgi\thanks{\tt georgi@physics.harvard.edu}\\ \\
  \small\sl Lyman Laboratory of Physics \\
  \small\sl Harvard University \\
  \small\sl Cambridge, MA 02138
}

\newsavebox{\moose}
\sbox{\moose}{%
\begin{picture}(0,0)
  \thicklines
  \put(-60,0){\circle{20}}
  \put(60,0){\circle{20}}
  \put(-50,0){\line(1,0){40}}
  \put(0,0){\circle{20}}
  \put(10,0){\line(1,0){40}}
  \put(-25,0){\vector(1,0){0}}
  \put(35,0){\vector(1,0){0}}
\end{picture}}
\newsavebox{\cmoose}
\sbox{\cmoose}{%
\begin{picture}(0,0)
  \thicklines
  \put(-60,0){\circle{20}}
  \put(60,0){\circle{20}}
  \dashline{6}(-50,0)(50,0)
  \put(0,0){\vector(1,0){0}}
\end{picture}}

\begin{document}
\begin{titlepage}
  \maketitle
  \begin{picture}(0,0)
    \put(400,200){\shortstack{
        \preprint HUTP-01/A024\\
        \preprint BUHEP-01-06\\
        \preprint UCB-PTH-01/15\\
        \rule{0pt}{8pt} }}
  \end{picture}
  
  \begin{abstract}
 
    We propose a new class of four-dimensional theories for natural
    electroweak symmetry breaking, relying neither on 
    supersymmetry nor on strong dynamics at the TeV scale.
    The new TeV physics is perturbative, and radiative 
    corrections to the Higgs mass are finite.  The softening of 
    this mass occurs because the Higgs is an extended object in theory
    space, resulting in an accidental symmetry. A novel Higgs 
    potential emerges naturally, requiring a second light $SU(2)$ 
    doublet scalar.

  \end{abstract}
  \thispagestyle{empty} \setcounter{page}{0}
\end{titlepage}

\section{Introduction}
\label{sec:introduction}

Experiments beginning later this decade will probe the fundamental
mystery of the weak interactions, the origin of electroweak symmetry
breaking. The standard model, a theory with a fundamental scalar field
that implements the Higgs mechanism, is almost certainly
incomplete---quadratically divergent radiative corrections to the
Higgs mass suggest that new physics is required at TeV energies to
stabilize the weak scale. To date theories for this stabilization can
be grouped into two categories: those which rely on new strong
dynamics or compositeness near a TeV (such as technicolor, composite
Higgs, or theories with a low fundamental Planck scale), and those
which are perturbative, using low-scale supersymmetry.  In this paper
we propose a third category: a class of non-supersymmetric theories
with a light Higgs, in which physics is entirely perturbative at the
TeV scale, yet the Higgs mass radiative corrections are finite.

Aside from supersymmetry there have been few ideas for ensuring a
light boson on purely symmetry grounds. Two directions are
spontaneously broken accidental global symmetries,
~\cite{Weinberg:1972fn} that can produce pseudo-Nambu-Goldstone
bosons, and gauge symmetries that protect vector boson masses. Neither
of these ideas seem directly relevant: the Higgs doesn't look much
like a Nambu-Goldstone boson (with non-derivative quartic, gauge, and
Yukawa couplings).\footnote{It is possible to make a
  pseudo-Nambu-Goldstone boson look like a Higgs by fine-tuning its
  mass term.  This is the idea behind composite Higgs models~\cite{
    Kaplan:1984fs}.}  Neither does it look like a boson with spin.
Consequently neither of these avenues has produced a viable theory
without the need for fine-tuning. Using the idea of dimensional
deconstruction~\cite{Arkani-Hamed:2001ca}, we will see that these two
ideas are in fact related to each other in a way which allows us to
construct realistic theories.  To motivate the connection, we briefly
detour into five-dimensions.

Consider an $SU(k)$ gauge theory in five dimensions compactified on a
circle of radius $R$. At distances large compared to $R$, the theory
appears four-dimensional, with an $SU(k)$ gauge symmetry.  The zero
mode of the five-dimensional gauge potential decomposes into the
four-dimensional $SU(k)$ gauge bosons and a real scalar field $\phi$
in the adjoint representation.  $\phi$ is associated with the
non-trivial Wilson loop $W = P \exp{(i \int dx_5 A_5)}$ around the
fifth dimension. Classically, $\phi$ is massless. At 1-loop,
$\phi$ picks up a quadratically divergent mass in the low-energy
four-dimensional effective theory.  However  at energies
much larger than $1/R$, $\phi$ is really $A_5$, the fifth component of
the gauge field. So, by the higher-dimensional gauge invariance, there
can't be any contribution to the mass of $\phi$ from energies much
higher than $1/R$, and therefore the quadratic divergence of the
low-energy theory must be cut-off at the scale $1/R$.  How is any
mass for $\phi$  generated?  Since the Wilson line is gauge
covariant, gauge invariance does not forbid the
operator $|\tr W|^2$, that contains a mass for the zero mode of $A_5$
in its expansion.  Since this is a non-local operator,
five-dimensional locality guarantees that it can not be generated with
a UV divergent coefficient. This is interesting, because the theory is
non-supersymmetric and even perturbative at the scale $1/R$, and yet
the correction to the $\phi$ scalar mass is completely finite.

In the next section we ``deconstruct'' this seemingly
higher-dimensional mechanism.  As we will see, the extra dimension is
not at all essential, and the physics can be understood in purely
four-dimensional terms. The light scalar is a pseudo-Nambu-Goldstone
boson, an object very familiar to gauge theory model builders.
Nevertheless, the higher dimensional picture will be a very useful
guide as we explore how this idea can be used as a starting point to
provide a new way to stabilize the Higgs mass in the standard
model\footnote{The idea that the Higgs might be a component of a
higher-dimensional gauge field in TeV-sized extra dimensions has a long
history; see for example \cite{extrad}.}.
There are many apparent obstacles to doing this. If the Higgs is to be
associated with components of a higher-dimensional gauge field, how
can we get it out of the adjoint representation? How can we get a
negative mass squared and a reasonably large quartic coupling, so the
Higgs gets a vev and the physical Higgs particle is sufficiently
heavy?

We will see that these issues are very naturally resolved in a 
six-dimensional theory. Non-adjoint scalars can be generated by 
enlarging the gauge group in the six dimensional bulk. Having special
sites where the gauge group is not enlarged allows us to give negative
mass squared to the Higgs. Finally, and perhaps most interesting,
the six-dimensional gauge kinetic energy contains non-derivative 
interactions which become a quartic coupling term between our
four-dimensional Higgs and another scalar doublet, stabilizing the
potential. 

If our theory were truly  six-dimensional, we would have the usual
higher-dimensional problems to contend with.  Why is the radius
stabilized near the TeV scale? What happens near the cutoff of the
higher-dimensional gauge theory? But in deconstruction, extra
dimensions are used purely as inspiration, and may be discarded at the
end, together with all the additional restrictions they imply. This
allows us to build realistic theories of electroweak symmetry breaking
in four dimensions with no higher-dimensional interpretation
whatsoever. The new feature of these theories is that the physics of
electroweak symmetry breaking remains perturbative and insensitive to
high-energy details up to a cut-off scale much larger than a TeV
without the need for any fine-tuning.

\section{Deconstruction}
\label{sec:5d-intu-deconstr}

We wish to deconstruct the toy 5D theory we have just described along the
lines of \cite{Arkani-Hamed:2001ca}.  Since the radiative stability of
the Higgs is a low energy problem, we may restrict our discussion to
the low energy effective Lagrangian of \cite{Arkani-Hamed:2001ca},
described by a ``condensed moose'' diagram. This corresponds to
putting the fifth dimension on a lattice with $N$ sites $i=1,\dots,N$
with periodic identification of site $i$ with site
$i+N$~\cite{Arkani-Hamed:2001ca,Hill:2000mu,Cheng:2001vd}.  On each
site there is an $SU(k)$ gauge group, and on the link pointing from
the $i$'th to the $(i+1)$'th site, there is a non-linear sigma model
field $U_i =$ exp$(i \pi_i^a T_a/f)$. Under the $SU(k)^N$ gauge
symmetry the link fields transform as $U_i \to \mathbf{g}_i U_i
\mathbf{g}_{i+1}^{-1}$. The effective Lagrangian is
\begin{equation}
  \label{eq:lattice}
  {\cal L} = -\frac{1}{2g^2} \sum_{i=1}^N
  \tr  F_i^2 + f^2
  \sum_{i=1}^N \tr \Bigl[(D_\mu U_{i})^\dagger D^\mu
  U_{i}\Bigr] +\dotsb 
\end{equation}
where the covariant derivative is $D_\mu U_{i} \equiv \partial_\mu
U_{i} -i A^i_\mu U_{i} + i U_{i} A^{i+1}_\mu$ and the dots represent
higher dimension operators that are irrelevant at low energies.%
\footnote{Similar models have been considered in other contexts,
  including models of CP violation \cite{Georgi:1974au} and quantum
  field theory final examinations \cite{Sidney:1993fe}.}  If the gauge
couplings are turned off, there is no coupling between the $U$s at
different sites and the theory has a large $SU(k)^{2N}$ accidental
``chiral'' global symmetry
\begin{equation}
  \label{eq:1}
  U_i \to L_i U_i R_{i+1}^{\dagger}  
\end{equation}
where $L_i,R_i$ are independent $SU(k)$ matrices. This is
spontaneously broken down to $SU(k)^N$, resulting in $N$ adjoint
Nambu-Goldstone boson multiplets.
 
The gauge couplings preserve only the $SU(k)^N$ subgroup of the global
symmetry \eqref{eq:1} where $L_i=R_i$.  Using the gauge freedom we can
almost go to a unitary gauge where all the $U_i$ are set to one. The
$SU(k)^N$ gauge theory is higgsed to the diagonal $SU(k)$ eating $N-1$
of the adjoint Nambu-Goldstone bosons along the way. The remaining
Nambu-Goldstone boson is associated with the product $U_1 U_2 \cdots
U_N$ which transforms homogeneously under $\mathbf{g}_1$ gauge
transformations and can not be gauged to unity.  This operator is the
discretization of the Wilson line in the continuum case. The linear
combination $\phi = (\pi_1 + \cdots \pi_N)/\sqrt{N}$ of
Nambu-Goldstone bosons is classically massless, and transforms as the
adjoint under the surviving diagonal $SU(k)$ gauge
group.\footnote{These objects were also noticed by
  \cite{Cheng:2001vd}.}  This field corresponds to 
the zero mode of $A_5$.  Because \eqref{eq:1} is broken by the gauge
interactions, $\phi$ gets a mass from loop effects.

We can characterize the breaking of \eqref{eq:1} through the
introduction of ``spurions'' $q_i$. We assign to $q_i$ a
transformation law so that the covariant derivative transforms
homogeneously. Taking
\begin{equation}
  \label{eq:2}
  D_\mu U_i = \partial_\mu U_i + i A_{\mu i} U_i - U_i q_{i} A_{\mu
    (i+1)}  q_{i}^\dagger 
\end{equation}
we can read off the transformation law
\begin{align}
  \label{eq:4}
  U_i &\to L_i U_i R_{i+1}^\dagger \\
  A_i &\to L_i A_i L_i^\dagger \\
  q_i &\to R_{i+1} q_i L_{i+1}^\dagger
\end{align}
There is also a separate symmetry $U(1)^N$ symmetry
under which
\begin{equation}
  \label{eq:5}
  q_i \to e^{i \alpha_i} q_i  
\end{equation}
Note that under a gauge transformation $U_i q_i\to L_i \, U_i q_i \,
L^\dagger_{i+1}$. After using $q_i$ to determine the symmetry
properties, we set $q_i=1$.

Now we can discuss the radiative corrections to the $\phi$ mass.  The
spurious symmetries tightly constrain the sorts of operators that can
be generated. The leading non-trivial operator involving only the $U$s
is
\begin{equation}
  \label{eq:6}
  {\cal O} =  |\tr (U_1 q_1 \cdots U_N q_N)|^2  
\end{equation}
The expansion of this operator gives a mass for $\phi$.  What do we expect
for the coefficient of this operator from matching to the UV theory? The
spurious symmetries along with standard power counting
\cite{Weinberg:1979kz, Manohar:1984md} yield a natural size for the
coefficient of this operator $(4\pi f)^2 f^2(g^2/16\pi^2)^N$. This gives a
$\phi$ mass suppressed by $\alpha^N$. For $N=1$ this is exactly the mass we
would expect from a quadratic divergence in the low energy theory with a
cut-off $4 \pi f$. However for $N>1$ this mass is significantly smaller
than the na\"\i{}ve one loop quadratic divergence would suggest.

In the low energy theory there are infrared contributions to the $\phi$
mass which can be much larger. At leading order in $g^2$ these come from
the Coleman-Weinberg potential. By the above power counting these
contributions are dominated by the infrared for $N>2$ and are therefore
insensitive to the cut-off and calculable in the low energy theory. 


Armed with this power-counting argument, let us calculate the leading
1-loop order low
energy contribution to the  $\phi$ potential. The analysis is most
efficiently done with the Coleman-Weinberg formalism. Turning on a
background value for $\phi$ corresponds to taking
\begin{equation}
  \label{eq:7}
  U_i = e^{i \phi/(f \sqrt{N})}  
\end{equation}
The $\sqrt{N}$ has been inserted to make $\phi$ a canonically
normalized field.

For simplicity, we consider the case of an $SU(k=2)$ theory. Then we
can always choose $\phi$ to point in the $\sigma^3$ direction, $\phi =
|\phi| \sigma^3$. The Coleman-Weinberg potential from gauge boson
loops is
\begin{equation}
  \label{eq:8}
  V(\phi) = \frac{3\Lambda^2}{32\pi^2} \tr M^2(\phi) + 
  \frac{3}{64\pi^2} \tr (M^2(\phi))^2 \log{\frac{M^2(\phi)}{\Lambda^2}}
\end{equation}
where $M^2(\phi)$ is the mass matrix for the $N$ gauge boson
multiplets in the presence of the background \eqref{eq:7} and
$\Lambda$ is the UV cut-off.  The discrete translation symmetry of the
theory allows diagonalization of the mass matrix by discrete Fourier
transform. The resulting eigenvalues for the charged gauge bosons are:
\begin{equation}
  \label{eq:9}
  m^2_n(\phi)   = 4 g^2 f^2 \sin^2\left(\frac{n\pi}{N} + \frac{\phi}{f \sqrt{N}}\right)
  \qquad -N/2<n \leq N/2
\end{equation}
Note that the sum of these eigenvalues is independent of $\phi$ for
$N>1$, and therefore the quadratic divergence in $V$ is a constant,
independent of $\phi$. For $N=2$, only a logarithmic divergence remains. 
The sum of the squares of these eigenvalues is
independent of $\phi$ for $N>2$, eliminating even the logarithmic
divergence. This explicitly confirms our expectations from
power-counting.  The (UV finite) potential for  $\phi$ for $N>2$
is~\cite{Georgi:1974au}
\begin{equation}
  \label{eq:10}
  V(\phi) = -\frac{9}{4 \pi^2} g^4 f^4 
\sum_{n=1}^{\infty} \frac{\cos(2 n \sqrt{N}
    \phi/f)}{n (n^2 N^2 - 1)(n^2 N^2 - 4)}+\text{constant}
\end{equation}
Here we see the absence of a divergence without cancellations between
particles of different statistics. Rather than eliminating the
potential entirely, as would happen in supersymmetry, the
cancellations here eliminate the \emph{dependence on $\phi$.}  That
is, the divergent renormalization is to the cosmological constant,
rather than the operator of \eqref{eq:6}. This cancellation is
guaranteed by our symmetry discussion; nevertheless it is amusing that
the contribution from any individual vector boson mass eigenstate is
divergent, while the spectrum and couplings are just right to ensure a
finite total result.

The spectrum of this theory exhibits an interesting hierarchy of
scales. The highest scale is the UV cut-off of the non-linear sigma
model $\Lambda = 4\pi f$. We refer to all the physics below this scale
as ``low energy.'' Physics above this scale can't be addressed within
the non-linear sigma model, although we can of course UV complete this
theory in a variety of (conventional) ways \cite{Arkani-Hamed:2001ca}:
``technicolor'', linear sigma model, supersymmetry, {\it etc.} The
$\phi$ mass is insensitive to the details of the physics at and above
$\Lambda$. Consequently we need not specify exactly what this physics
is for our purposes.  Below $\Lambda$ we have a tower of massive
vector bosons, extending down in mass from $\sim g f$ to $\sim g f/N$.
For large $N$ the states near the bottom reproduce the spectrum of an
extra-dimensional gauge theory compactified on a circle
\cite{Arkani-Hamed:2001ca}, while for small $N$ (say 3), no
extra-dimensional interpretation is possible. Next, the light scalar
$\phi$ has a mass squared $m^2_\phi \sim g^4 f^2/(16\pi^2 N^3)$. The
gauge coupling of the unbroken $SU(2)$ is $g/\sqrt{N}$, so this mass
squared is a loop factor smaller than the lightest massive vector
boson mass squared, $\sim g^2 f^2/N^2$. Finally we have a massless
$SU(2)$ gauge boson.  We refer to physics near or below the
$\phi$ mass as ``very low energy.'' In the very low energy theory at
this point, we have only the $\phi$ and the massless gauge bosons.

Note that the $\phi$ mass is what we would expect from the apparent
quadratic divergence in the very low energy theory but with a cutoff
of only $\sim g f/N$. This is the scale of new physics: the bottom of
the tower of massive vector bosons.  This physics is entirely
perturbative: no strong interactions are required at this scale to cut
off the quadratic divergence. This is to be contrasted with the
expectation from pseudo-Nambu-Goldstone bosons in QCD or technicolor.
For instance, in the limit where the quark masses vanish in QCD, there
is a quadratically divergent contribution to the charged pion mass
from photon loops in the low-energy pion effective theory.  The pion
mass is then of order $e/4\pi$ times $\Lambda \sim 4 \pi f_\pi$, the
cutoff of the pion effective theory. On the other hand, in our
effective theory, there are {\it no} divergent contributions, the
Higgs mass we compute is insensitive to the detailed physics at $4 \pi
f$, and the result is smaller by an additional factor of
$g/(4\pi\sqrt{N})$. In the full theory, this is obvious, as the
accidental symmetry prevents the appearance of a counterterm that
could absorb the quadratic divergence. But in the very low energy
theory, the lightness of $\phi$ looks miraculous.  It is this that
makes $\phi$ an interesting starting point for a model of the Higgs.

In our realistic models the scale of new perturbative physics will be
near 1 TeV, while the scale where the non-linear sigma model
description breaks down will be parametrically larger, numerically
between 10 and 100 TeV. This is similar to what happens in composite
Higgs models~\cite{Kaplan:1984fs}, but the difference here is that we
will not require any fine tuning to maintain this ratio of scales.

What was essential for this mechanism to work? The light scalar
descended from a ``chain'' of non-linear sigma model fields, a
``non-local'' object in the space of gauge theories. Since the gauge
interactions themselves \emph{are} local in this space, the only
operator that gives rise to the scalar mass must involve the whole
chain. This operator is then of very high scaling dimension, and is
generated with a finite coefficient. We refer to naturally light
scalars of this kind as ``chain scalars''.

\section{Realistic Theories}
\label{sec:realistic-theories}

In order to describe the Higgs field in the Standard Model we need a
scalar transforming as a doublet under $SU(2)$, rather than in the
adjoint representation.  Furthermore we need a potential that breaks
the electroweak symmetry and leaves the physical Higgs scalar somewhat
heavier than the $Z$. This requires a negative mass squared and a
substantial quartic self-coupling for the Higgs field.  Finally we
need an order one Yukawa coupling to the top quark.  We need to
incorporate all these features without reintroducing a quadratic
divergence for the $\phi$ mass squared.

Let us first try to get our ``chain'' scalars out of the adjoint
representation of the low-energy gauge group. This can happen if the
theory distinguishes different components of the adjoint. This is only
possible if the low-energy gauge group is reduced. Consider, for
instance, a condensed moose diagram with $SU(3)$ gauge symmetries on
the sites $i=2,\cdots,N$, with only the $SU(2) \times U(1)$ subgroup
of $SU(3)$ leaving $T_8$ invariant at $i=1$.  All the link variables
continue to be $3 \times 3$ special unitary matrices, but the matrix
$\mathbf{g}_1$ resides only in the $SU(2) \times U(1)$ direction, so
that $\mathbf{g}_1$ commutes with $T_8$.  In the continuum 5D picture,
this is a five-dimensional theory with an $SU(3)$ gauge symmetry in
the bulk, together with a ``brane'' where the gauge symmetry is
reduced to $SU(2) \times U(1)$.

The fluctuations of the $U$s now higgs the theory down to $SU(2)
\times U(1)$. The 8 components of the chain scalar decompose under
this $SU(2) \times U(1)$ as $\text{\bf 3}_0 \oplus \text{\bf 1}_0
\oplus \text{\bf 2}_{1/2} \oplus \text{\bf 2}_{-1/2}$.  (This
normalization of the $U(1)$ corresponds to the decomposition
$\text{\bf 3} \to \text{\bf 2}_{1/6} \oplus \text{\bf 1}_{-1/3}$.)
The last two have the quantum numbers of the standard model Higgs
field and its complex conjugate.  Since the gauge symmetry no longer
relates these different components, they can pick up different masses.
More explicitly, the reduced gauge symmetry on the first site allows
additional operators in the theory. Since $\mathbf{g}_1$
commutes with $T_8$, the link variable $U_1$ and $T_8 U_1$ have
identical transformation properties under all symmetries, as do $U_N$
and $U_N T_8$.

As emphasized in \cite{Kaplan:1984fs}, to build a Higgs, it is
not enough to ensure that it has a small mass. It must also have
quartic self-interactions that are large compared to its mass squared
over the cut-off squared. The Coleman-Weinberg interactions that
produce a small $\phi$ mass also produce quartic interactions, but
these are suppressed by a similar factor.  Thus we also need some
other source for a quartic potential for the Higgs fields. But the
additional self-interactions must not disturb the crucial cancellation
of quadratic divergences. Again we can take inspiration from
higher-dimensional physics, in which the $\phi$ is related to a gauge
field. Because gauge boson self-interactions contain non-derivative
terms, we should be able to build non-derivative interactions for the
$\phi$. Suppose, for example, that we start in 6D with an $SU(3)$
gauge theory.  Then the action contains a piece $\tr F_{56}^2$, that
yields a quartic potential
\begin{equation}
  \label{eq:10a}
  \tr([A_5,A_6]^2)
\end{equation}
for the zero modes of $A_5,A_6$ in the low-energy theory. Because of
the higher dimensional gauge invariance, this should not introduce any
divergent masses for the zero modes.

What is the analog of this operator in the condensed moose language?
Consider a condensed moose diagram that is the discretization of a
torus with $N \times N$ sites, labeled by integers $(i,j)$. The sites
$i$ and $i+N$ are identified, as are $j$ and $j+N$. We also have link
fields $U_{i,j}$ between the sites $(i,j)$ and $(i,j+1)$, and
$V_{i,j}$ between $(i,j)$ and $(i+1,j)$. Finally we add the ``plaquette''
operators to the action:
\begin{equation}
  \label{eq:11}
  -\sum_{i,j} \lambda_{i,j} f^4 \tr (U_{i,j} V_{i,j+1} U^\dagger_{i+1,j}
  V^\dagger_{i,j}) + \text{h.c.}
\end{equation}
Some of the Nambu-Goldstone bosons are eaten in higgsing the gauge
group down to the diagonal $SU(3)$. Others become massive through the
plaquette potential. For any values of the $\lambda_{i,j}$ two
massless multiplets remain, the analog of the continuum zero modes. It
is easy to check that these modes correspond to
\begin{align}
  \label{eq:12}
  U_{i,j} & = e^{i u/(f N)} \\
  V_{i,j} & = e^{i v/(f N)}
\end{align}
{\it i.e.} uniform link variables in the two directions.
With this normalization, $u,v$ are canonically normalized fields.  
For these modes the plaquette action expands to the quartic potential
\begin{equation}
  \label{eq:13}
  \mbox{constant} + \frac{\lambda}{N^2} \tr |[u,v]|^2 + \cdots, \qquad 
   \lambda \equiv \frac{1}{N^2}\sum_{i,j}  \text{Re}\,\lambda_{i,j}
\end{equation}
As expected from the gauge theory analog, 
the spurious symmetries of the theory are enough to guarantee
the absence of divergences for the radiative corrections to these scalar 
masses for $N>2$, and only logarithmic divergences for $N=2$.

Note that the form of the self-interactions in \eqref{eq:13} is the 
trace of the square of a commutator. This special form is required 
to avoid quadratic divergences, and it is an obvious reminder of
the analog, \eqref{eq:10a}, in the six-dimensional gauge theory.
In the very low energy theory additional couplings will be generated
through low-energy renormalization.

In order to get the Higgs out of the adjoint representation, we
replace the gauge group at one of the sites, say $(1,1)$, with $SU(2)
\times U(1)$.  It is easy to show that the classically massless scalar
multiplets continue to be those of the form \eqref{eq:12}.  As
before the reduced gauge symmetry allows additional operators, in this
case involving an insertion of $T_8$ in the four plaquette terms
that touch the $(1,1)$ site. For instance
\begin{equation}
  \label{eq:15}
  (\alpha + i \beta) f^4 \tr T_8 U_{1,1} V_{1,2} U^\dagger_{2,1}
  V^\dagger_{1,1}  +\text{h.c.} 
\end{equation}
For the zero modes \eqref{eq:12} this becomes a mass term
\begin{equation}
  \label{eq:16}
    - i \beta \frac{f^2}{N^2} \tr T_8 [u,v] + \cdots
\end{equation}
If this operator is included with a large coefficient, then there is a
large tree-level mass term for the scalars in the theory. However,
note that this operator is odd under the interchange $U_{i,j}
\leftrightarrow V_{j,i}$.  If we impose the symmetry $U_{i,j}
\leftrightarrow V_{j,i}$ on the theory, then it is technically natural
for $\beta$ to be small compared to the $\lambda_{i,j}$.

We now have all the ingredients to construct a realistic theory of
electroweak symmetry breaking.  It is convenient to group $u,v$ into
the matrix
\begin{equation}
  {\cal H} = \frac{u + i v}{\sqrt{2}} = 
  \begin{pmatrix} \varphi + \eta &
    h_1 \\ h_2^{\dagger} & -2 \eta 
  \end{pmatrix}
\end{equation}
where $\varphi,\eta$ are complex fields in the $\text{\bf 3}_0,
\text{\bf 1}_0$ representation of $SU(2) \times U(1)$ respectively,
and $h_1, h_2$ have the quantum numbers $\text{\bf 2}_{1/2}$ of the
standard model Higgs.  The quartic potential is
\begin{equation}
  \label{eq:16b}
  \lambda \tr [{\cal H},{\cal H}^{\dagger}]^2 = \lambda 
  \tr (h_1 h_1^{\dagger} - h_2 h_2^{\dagger})^2
  + \lambda (h_1^{\dagger} h_1 - h_2^{\dagger} h_2)^2 + \text{terms
    involving}  \, \varphi, \eta.
\end{equation}
Note the similarity of this quartic potential to the one in the
supersymmetric standard model. Here, it follows from the special form
\eqref{eq:13}.

The Coleman-Weinberg potential generates positive mass squared for all
of $\varphi,\eta, h_1,h_2$. We can also add other operators, such as
$\tr(U_{2,1} U_{2,2} \cdots U_{2,N}) + \text{h.c.} + (U_{i,j} \to
V_{j,i})$, with small coefficients (since these break the spurious
global symmetries it is technically natural for their coefficients to
be small).  These are of dimension $N$ and therefore do not affect our
power counting analysis for the finiteness of the Higgs mass for
$N<5$; and they also preserve the $U \leftrightarrow V$ symmetry. They
give rise to the same positive squared mass for all of
$h_1,h_2,\varphi,\eta$.  In any case, in order to obtain a realistic
theory, we need to distinguish between $h_1$ and $h_2$. The reason is
familiar from similar considerations in the supersymmetric standard
model: since the quartic potential for $h_1, h_2$ has a flat direction
where $|h_1| = |h_2|$, we must have $m_{h_1}^2 + m_{h_2}^2 > 0$ in
order not to run away along this flat direction. If the symmetry
between $h_1, h_2$ is unbroken, this forces both $m^2$s to be positive
and there can be no electroweak breaking. Fortunately, the operator in
\eqref{eq:16} distinguishes between $h_1, h_2$. In fact
\begin{equation}
  \label{eq:16a}
  -i \beta \frac{f^2}{N^2} \tr T_8 [u,v] =  \beta \frac{f^2}{N^2}
  \tr T_8 [{\cal H}, {\cal H}^\dagger] 
  = \beta \frac{f^2}{N^2} (h_2^\dagger
  h_2 - h_1^\dagger h_1)
\end{equation}

This operator can make one of the masses, say $m_{h_1}^2$, negative,
while keeping all the others positive. As we have seen, there is also
an $O(1)$ quartic Higgs coupling. To give rise to a small Higgs vev,
the coefficient $\beta$ in \eqref{eq:16} must be chosen so that this
contribution to the $h$ masses is the same order of magnitude as the
Coleman-Weinberg and other contributions. Thus the masses of $h_2$ and
the rest of the ${\cal H}$ multiplet are expected to be of the same
order of magnitude as the Higgs. But no special fine tuning is
required---all these masses are safe from quadratic divergences. There
is then a wide range of parameters for which $SU(2) \times U(1)$ is
broken in the correct way by the vacuum expectation value of
$h_1$.\footnote{Other terms in \eqref{eq:15} besides \eqref{eq:16}
  have interesting phenomenological consequences, for example, giving
  rise to a small vev for $\varphi$ and an (acceptably small)
  contribution to the $T$ 
  parameter.}

We have succeeded in triggering electroweak symmetry breaking in a
natural way. What aspect of the phenomenology of this model can be
used to check our mechanism for stabilizing the electroweak hierarchy?
In the very low energy theory, below a TeV, we will see not only the
Higgs, but the other pseudo-Nambu-Goldstone bosons. The precise
spectrum is model dependent, but the existence of a light $h_2$ and
the specific quartic interaction \eqref{eq:16b} is a robust prediction
of this mechanism. To make the Higgs much lighter than the $h_2$
multiplet requires a fine-tuning. At TeV and higher energies, we must
see the non-linear sigma model structure with local couplings in
theory space. This would reveal itself through low-energy theorems for
the scattering of the Nambu-Goldstone bosons, in this case the Higgs
fields, $h_2$ and the longitudinal components of the additional vector
bosons at multi-TeV energies.

\section{Fermions}
For a realistic model, we need fermions with Yukawa couplings to the
Higgs.
Like gauge couplings and quartic Higgs self-couplings, large Yukawa
couplings generically induce a quadratically divergent mass squared
for the Higgs boson in the low-energy theory. But this divergence can
also be avoided using ``locality'' in theory space.  Before
constructing the appropriate local interactions, let us ask what the
analogue of the Higgs Yukawa coupling looks like in our language.  As
a preliminary example, consider the one-dimensional chain with $SU(3)$
gauge symmetries on the sites $i=2,\cdots ,N$ and $SU(2) \times U(1)$
at $i=1$.  We also introduce the standard model fermions with their
usual quantum numbers under $SU(3)_{\text{color}} \times SU(2) \times
U(1)$: $Q \sim (\text{\bf 3,2})_{1/6}, U^c \sim (\mathbf{\bar{3}}
,\text{\bf 1})_{-2/3}, D^c \sim (\mathbf{\bar{3}} ,\text{\bf
  1})_{+1/3}, L \sim ({\text{\bf 1,2})_{-1/2}, E^c \sim (\text{\bf 1,
    1}})_{+1}$. The Yukawa couplings to the chain ``Higgs'' field can
arise from the gauge invariant operator
\begin{equation}
\label{eq:yuk1}
\begin{pmatrix} Q & 0 \end{pmatrix} U_1 \dots U_N \begin{pmatrix} 
0 \\ 0 \\ U^c \end{pmatrix}
\end{equation}
for the up Yukawa couplings, and 
\begin{equation}
\label{eq:yuk2}
\begin{pmatrix} 0 & 0 & D^c \end{pmatrix} U_1 \dots U_N \begin{pmatrix} 
Q\\ 0 \end{pmatrix},\quad \begin{pmatrix} 0 & 0 & E^c 
\end{pmatrix} U_1 \dots U_N \begin{pmatrix} 
L \\ 0 \end{pmatrix}
\end{equation}
for the down and charged lepton Yukawa couplings. 
        
However, because these operators correspond to ``non-local''
interactions, adding them to the theory would give rise to a quadratic
divergence for the Higgs mass. This may not be a problem for the
Yukawa couplings for the light generations, since the coefficient of
the quadratic divergence is then easily small enough so that cutting
it off at the non-linear sigma model scale does not disturb the light
Higgs. However, this is not the case for the top Yukawa coupling. It
is technically natural to eliminate these operators, but then we lose
the Yukawa coupling to the fermions.  What we would like to do instead
is obtain these effective operators in the very low-energy theory,
starting with purely local interactions in theory space. This can be
done in a by now familiar way. For simplicity we consider only the top
Yukawa coupling. We  add vector-like fermions
$\chi_i,\chi_i^c$ for $i=2,\cdots,N$, that transform as triplets and
anti-triplets under $SU(3)_i$, anti-triplets and triplets under
$SU(3)_{\text{color}}$, and have $U(1)$ charges $\mp 1/3$.
Note that these fermions now transform under gauge interactions other
than the $SU(3)_i$ at their site.  Together with mass terms, these
$\chi$s have nearest-neighbor couplings through the $U$s:
\begin{equation}
y_1 f \begin{pmatrix} Q & 0 \\ \end{pmatrix} U_1 \chi_2 + \sum_{i=2}^{N-1}
\chi^c_i(y_i f \chi_i - y'_i f U_i \chi_{i+1}) + y_N f \chi^c_N U_N
\begin{pmatrix} 0 \\ 0 \\ U^c \end{pmatrix}
\end{equation}
Integrating out the massive $\chi,\chi^c$ leaves the massless fields
$Q,U^c$ coupled to the Higgs in the low-energy theory. However, with
these local interactions, our power-counting analysis guarantees the
absence of all divergences in the Higgs mass for $N>2$, and only
logarithmic divergences for $N=2$.

Finally, we want to introduce fermions in the $N \times N$ site model
of the previous section, where the Higgs triggers electroweak symmetry
breaking.  This can be done in a number of ways. For instance, we can
have a chain of fermions as in the previous paragraph, in both the $U$
and $V$ directions symmetrically, to preserve the $U \leftrightarrow
V$ symmetry.

Note that the operators \eqref{eq:yuk1},\eqref{eq:yuk2} may be used to
generate the Yukawa couplings for the 2 light generations. In this
case the only interactions which break the $U(2)^5$ flavor symmetry
are the Yukawa couplings themselves. This guarantees the absence of
dangerous flavor changing neutral currents via the GIM mechanism.

\section{Conclusions}
\label{sec:conclusions}

We have seen that if the Higgs field descends from a chain of links in
theory space, we can trigger electroweak symmetry breaking in a way
that remains perturbative and insensitive to high-energy details up to
a cut-off scale much larger than a TeV without the need for any
fine-tuning. The radiative corrections to the Higgs mass are
controllably small, without relying on supersymmetry or strong
dynamics at the TeV scale.  Notice in particular we have included, in
a natural way, a set of operators with varying sizes. A fundamental
theory above the scale where our non-linear sigma model description
breaks down will determine which operators appear in the low energy
theory. However, we have seen that it is natural to allow some
``local'' operators with large coefficients, while higher-dimensional
``non-local'' operators have small ones.

But where does ``locality'' come from? Normally, locality of
interactions in position space is simply taken for granted. However,
as we emphasized in \cite{Arkani-Hamed:2001ca}, there is an intimate
connection between physical space and theory space, where locality can
have a deeper origin. For instance in the simple constructions of
\cite{Arkani-Hamed:2001ca}, higher-dimensional locality is a
consequence of the renormalizability of the 4D gauge theory that
dynamically generated the extra dimension. We expect that similar
considerations will provide a deeper explanation of the pattern of
local operators in our present constructions, once our non-linear
sigma model is UV completed in a renormalizable theory at high
energies, above $\sim 10$ to 100 TeV.

Our models provide a realistic theory of
electroweak symmetry breaking, although
they are by no means the unique implementation of
our central mechanism. For instance, elegant ``orbifold'' theories can be
constructed where the only light scalars are $SU(2)$ doublets. But it
is crucial that there be a light $h_2$ partner of the Higgs doublet with
the squared commutator self-interaction of \eqref{eq:16b}.

The theories we have constructed have a rich and novel phenomenology
at TeV energies. In addition to a light Higgs, they have a distinctive
spectrum of new scalar, fermionic and vector particles with
perturbative couplings (for moderate $N$).  In this decade, experiment
will help us determine whether or not extended objects in theory space
are relevant to unraveling the mystery of electroweak symmetry
breaking.

\section{Acknowledgments}
N.A-H. thanks Lawrence Hall for valuable suggestions on obtaining the
realistic standard model group structure. We are grateful to David Kaplan
for clarifying discussions on power-counting. H.G. is supported in part by the National
Science Foundation under grant number NSF-PHY/98-02709. A.G.C. is supported
in part by the Department of Energy under grant number
\#DE-FG02-91ER-40676.  N.A-H. is supported in part by the Department of
Energy. under Contracts DE-AC03-76SF00098, the National Science Foundation
under grant PHY-95-14797, the Alfred P. Sloan foundation, and the David and
Lucille Packard Foundation.


\begin{thebibliography}{1}

\bibitem{Weinberg:1972fn}
S.~Weinberg, ``Approximate symmetries and pseudoGoldstone bosons,'' {\em Phys.
  Rev. Lett.} {\bf 29} (1972) 1698--1701.

\bibitem{Kaplan:1984fs}
D.~B. Kaplan and H.~Georgi, ``SU(2) x U(1) breaking by vacuum misalignment,''
  {\em Phys. Lett.} {\bf B136} (1984) 183; S. Dimopoulos {\it et. al.},
``Composite Higgs scalars'', {\em Phys. Lett.} {\bf B136} (1984) 187.

\bibitem{Arkani-Hamed:2001ca}
N.~Arkani-Hamed, A.~G. Cohen, and H.~Georgi, ``(De)constructing dimensions,''
  \href{http://xxx.lanl.gov/abs/hep-th/0104005}{{\tt hep-th/0104005}}.

\bibitem{extrad} 
N.~S.~Manton,
``A New Six-Dimensional Approach To The Weinberg-Salam Model,''
{\em Nucl.\ Phys.}  {\bf B158}, (1979) 141; N.~V.~Krasnikov,
``Ultraviolet fixed point behavior of the five-dimensional Yang-Mills
theory, the gauge hierarchy problem and a possible new dimension at the 
TeV scale,'' {\em Phys.\ Lett.} {\bf B273}, (1991) 246;
G.~Dvali, S.~Randjbar-Daemi and R.~Tabbash,
``The origin of spontaneous symmetry breaking in theories with large  extra dimensions,''
\href{http://xxx.lanl.gov/abs/hep-ph/0102307}{{\tt hep-ph/0102307}}.


\bibitem{Hill:2000mu}
C.~T. Hill, S.~Pokorski, and J.~Wang, ``Gauge invariant effective Lagrangian
  for Kaluza-Klein modes,'' \href{http://xxx.lanl.gov/abs/hep-th/0104035}{{\tt
  hep-th/0104035}}.

\bibitem{Cheng:2001vd}
H.-C. Cheng, C.~T. Hill, S.~Pokorski, and J.~Wang, ``The standard model in the
  latticized bulk,'' \href{http://xxx.lanl.gov/abs/hep-th/0104179}{{\tt
  hep-th/0104179}}.

\bibitem{Georgi:1974au}
H.~Georgi and A.~Pais, ``CP - violation as a quantum effect,'' {\em Phys. Rev.}
  {\bf D10} (1974) 1246.

\bibitem{Sidney:1993fe}
S.~Coleman, ``Quantum Field Theory Final Examination,'' 1997.

\bibitem{Weinberg:1979kz}
S.~Weinberg, ``Phenomenological Lagrangians,'' {\em Physica} {\bf A96} (1979)
  327.

\bibitem{Manohar:1984md}
A.~Manohar and H.~Georgi, ``Chiral quarks and the non-relativistic quark
  model,'' {\em Nucl. Phys.} {\bf B234} (1984) 189.

\end{thebibliography}

\providecommand{\href}[2]{#2}\begingroup\raggedright\endgroup

\end{document}